\documentstyle[aps,epsfig]{revtex}
\begin{document}
\draft
\tighten
\title{Phase transition in linear sigma model and disoriented
chiral condensate}
\author{\bf A. K. Chaudhuri \cite{*}}
\address{
Variable Energy Cyclotron Centre\\
1-AF, Bidhan Nagar, Calcutta- 700 064}
\maketitle
\begin{abstract}
We  have investigated the phase transition and disoriented chiral
condensate domain formation in linear sigma  model.  Solving  the
equation  of  motion for the sigma model fields in contact with a
heat bath, we have shown that the fields undergo phase transition
above a certain temperature ($T_c$). It was also shown that  when
the fields thermalised at temperature above $T_c$ are cooled down
sufficiently  rapidly,  disoriented chiral condensate domains are
formed quite late in the evolution.
\end{abstract}


The possibility of forming disoriented chiral condensate (DCC) in
relativistic  heavy  ion  collisions  has  generated considerable
research activities in recent years. The idea was first  proposed
by Rajagopal and Wilczek \cite{ra93,ra93a,ra95,wi93}. They argued
that  for  a  second  order  chiral  phase transition, the chiral
condensate   can   become   temporarily   disoriented   in    the
nonequilibrium conditions encountered in heavy ion collisions. As
the  temperature drops below $T_c$, the chiral symmetry begins to
break  by  developing  domains  in  which  the  chiral  field  is
misaligned  from its true vacuum value. The misaligned condensate
has the same quark content and quantum numbers as  do  pions  and
essentially  constitute  a  classical pion field. The system will
finally relaxes to the true vacuum and in the  process  can  emit
coherent  pions.  Since the disoriented domains have well defined
isospin orientation,  the  associated  pions  can  exhibit  novel
centauro-like  \cite{la80,ar83,al86,re88} fluctuations of neutral
and charged pions \cite{an89,an89a,bj93,bl92}.

Most dynamical studies of DCC have been based on the linear sigma
model,  in  which  the chiral degrees of freedom are described by
the real O(4)  field  $\Phi=(\sigma,\roarrow{\pi})$, with the
Lagrangian,

\begin{equation}
{\mathcal{L}} =\frac{1}{2} (\partial_\mu \Phi)^2 -
\frac{\lambda}{4}
(\Phi^2 - f_\pi^2)^2, \label{0}
\end{equation}

\noindent  where  $\lambda$  is  a positive coupling constant and
$f_\pi$ is the pion decay constant.  At  finite  temperature,  to
leading  order  in  $\lambda$,  the thermal fluctuations $<\delta
\phi^2>$  of  the  pions  and  $\sigma$-mesons  do  generate   an
effective Hartree type dynamical mass giving rise to an effective
temperature  dependent  potential  $\lambda  T^2/2$  \cite{bo96}.
Resulting  chiral  phase  transition  is  compatible   with   the
expectations  of  lattice gauge QCD calculations \cite{go91}. One
generally introduces a finite symmetry breaking  term  $h_\sigma$
to  take  into account the finite pion mass. However, at present,
we are ignoring such terms, as our  interest  is  to  investigate
chiral   phase   transition  and  subsequent  disoriented  chiral
condensate domain formation. With symmetry breaking  term,  there
is no exact phase transition.

Aim  of  the present letter is to study the influence of external
source  on  the  phase  transition  and  subsequent  DCC   domain
formation  when  the  chiral  symmetric  phase  relaxes  back  to
symmetry broken phase. Indeed, in a heavy ion collision, while it
is possible that in a certain region chiral symmetry is restored,
that region  must  be  in  contact  with  some  environment  i.e.
background.  Exact  nature  of  the  environment  is difficult to
determine but presumably it will be consists of mesons and hadron
(pions, nucleon etc.). Recognising the uncertainty in  the  exact
nature  of  the environment, we choose to represent it by a white
noise  source,  i.e.  a  heat  bath.  To   be   consistent   with
fluctuation-dissipation  theorem,  we  also include a dissipative
term in the equation of motion for the sigma model  fields.  Thus
we  are  studying  essentially  Langevin  equation  for  the O(4)
fields. Recently it has been shown that in the  $\phi^4$  theory,
hard modes can be integrated out on a two loop basis resulting in
a  Langevin  type  equations for the soft modes \cite{gr97,xu99},
there by justifying our  approach.  Langevin  equation  for  O(4)
fields  have  been  used  by several authors \cite{bi97,ch99a} to
study the interplay of friction and white  noise  in  disoriented
chiral  condensate  formation. We thus propose to study following
Langevin equation,

\begin{equation}
[\frac{\partial^2}{\partial \tau^2} +(\frac{1}{\tau}+\eta)
\frac{\partial}{\partial \tau}
-\frac{\partial^2}{\partial x^2} -\frac{\partial^2}{\partial y^2} -
\frac{1}{\tau^2} \frac{\partial^2}{\partial Y^2}
+\lambda (\Phi^2 - f^2_\pi -T^2/2)] \Phi
 = \zeta (\tau ,x,y,Y)\label{1}
\end{equation}

\noindent  where  we  have  used proper time ($\tau$) and rapidity
(Y),  which  are  the  appropriate  coordinates  for  heavy   ion
scattering.  In  eq.  \ref{1}  $\eta$  is  the  friction. As told
earlier,  the  environment  or  the  heat  bath   ($\zeta$)   was
represented  as  a  white  noise  source  with  zero  average and
correlation as demanded by fluctuation-dissipation theorem,

\begin{mathletters}
\begin{eqnarray}
<\zeta(\tau,x,y,Y)> =&&0\\
\int <\zeta_a(\tau_1,x_1,y_1,Y_1)\zeta_b(\tau_2,x_2,y_2,Y_2)> d^4x
=&& 2 T \eta \delta_{ab}
\end{eqnarray}
\label{1a}
\end{mathletters}

Set  of partial differential eqs. \ref{1} were solved on a $32^3$
lattice using a lattice spacing of 1 fm, thus cutting  off  modes
with  momenta  $>$200 MeV. To show that the model undergoes phase
transition we keep the heat bath at  fixed  temperature  $T$  and
evolve  the  fields  for sufficiently long time (30 fm) such that
equilibrium is reached. Fields thermalised at higher  temperature
will   be  more  randomized  than  fields  thermalised  at  lower
temperature. At low temperature, the randomisation  will  not  be
complete  and $<\sigma>$ will have definite non-zero value. Above
the critical temperature ($T_c$), randomisation will be  complete
and $<\sigma>$ will be zero, indicating restoration of symmetry.

Solving   eqn.\ref{1}  require  initial  conditions  ($\phi$  and
$\dot{\phi}$). We distribute the initial fields  according  to  a
random Gaussian with,

\begin{mathletters}
\begin{eqnarray}
<\sigma>=&&f(r)f_\pi \\
<\pi_i>=&&0 \\
<\sigma^2>-<\sigma>^2 = <\pi_i^2>-<\pi_i>^2=   && f_\pi^2/4 f(r)\\
< \dot{\sigma}>=&& <\dot{\pi_i}>=0\\
<\dot{\sigma}^2>=<\dot{\pi}>^2=&& f_\pi^2 f(r)
\end{eqnarray}
\label{2}
\end{mathletters}

The interpolation function

\begin{equation}
f(r)=[1+exp(r-r_0)/\Gamma)]^{-1}
\end{equation}

\noindent  separates  the  central  region  from  the rest of the
system. We have  taken  $r_0$=11  fm  and  $\Gamma$=0.5  fm.  The
initial field configurations corresponds to quench like condition
\cite{ra93}   but   it  is  important  to  note  that  the  field
configuration at  equilibrium  will  be  independent  of  initial
configuration. Equilibrium value depend on the heat bath only. We
have  verified this in our code. The other parameter of the model
is the friction ($\eta$). In the present paper, we  use  $\eta  =
\eta_\pi  +\eta_\sigma$  and for $\eta_\pi$ and $\eta_\sigma$ use
values as calculated by Rischke \cite{ri98} but once  again,  its
precise  value  is  not of importance here, as we are looking for
fields at equilibrium. Friction determines the rate  of  approach
to equilibrium. This aspect of equilibration was also verified.

The  condensate  value of the $\sigma$ field can be considered as
the order parameter for the phase transition. If there is a phase
transition at $T_c$, the order  parameter  should  {\em  exactly}
vanish  for  temperatures  $\geq$ $T_c$, while below $T_c$ it will
have  nonzero  values  \cite{landau}.  We  calculate  the   order
parameter at the end of the evolution as,

\begin{equation}
<\sigma> = \frac{\int \sigma dx dy dY}{\int dx dy dY} .
\end{equation}

In fig.1, we have shown the variation of the order parameter with
temperature.  At low temperature, as expected, order parameter is
around $f_\pi$=92 MeV. With  increasing  temperature,  $<\sigma>$
decreases  and  become  zero  around  $T_c$=120  MeV,  indicating
restoration  of  chiral  symmetry.  It  remain  zero  at   higher
temperatures  also.  The  behaviour  of  the  $\sigma$ condensate
corresponds to a second  order  phase  transition.  The  critical
temperature  $T_c$=120  MeV  is also in agreement with mean field
calculations.

Chirally  symmetric  phase  at high temperature will roll back to
symmetry broken phase as the system cools and  temperature  drops
below the transition temperature. As told earlier, it is has been
conjectured  \cite{ra93} that domain like structure with definite
isospin content may emerge during the roll down period. Numerical
simulations of linear sigma model indicate that with quench  like
initial condition, domain like configurations do indeed emerge as
the  chirally  symmetric phase roll back to symmetry broken state
\cite{as94}. However, in heavy ion collision,  quench  is  not  a
natural  initial  condition. $<\phi>$ and $<\dot{\phi}>$ are in a
configuration  appropriate  for  high  temperature  but  that  of
$<\phi^2>$  and  $<\dot{\phi}^2>$  are  characteristic of a lower
temperature. On  the  contrary,  thermalised  fields  are  better
suited  to  mimic  initial conditions that may arise in heavy ion
collisions. Here, $<\phi>$, $<\dot{\phi}>$ as well as  $<\phi^2>$
and  $<\dot{\phi}^2>$ are in a configuration appropriate for high
temperature.

To  see  whether  domain  like structure emerges or not with more
appropriate initial condition like the thermalised fields, we did
a demonstrative calculation. The fields were thermalised at T=200
MeV. We then allow the heat bath to cool and follow the evolution
of the thermalised fields. Evolution of  the  thermalised  fields
now  will  depend on the exact nature of the friction, however we
choose to  use  the  same  friction  as  before.  We  assume  the
following cooling law for the heat bath,

\begin{equation}
T(t)=T_0 \frac{1}{t^n}
\label{4}
\end{equation}

\noindent  with  n=1  (appropriate  for  3d  scaling  expansion).
Assuming that the number density is proportional to the square of
the amplitude, at each lattice point, we calculate the neutral to
total pion ratio according to,

\begin{equation}
R(x,y,Y) =\frac{N_{\pi_0}}{N_{\pi_1} + N_{\pi_2} +N_{\pi_0}}
\end{equation}

Very  large or small value of the ratio, over an extended spatial
zone will be definite indication of disoriented chiral condensate
domain formation. At this  point  we  would  like  to  note  that
contour  plot  of  one  single component of the pion field as has
been shown in ref. \cite{as94,ch99b} donot  necessarily  indicate
domain formation.

In fig.2 we have shown contour plot of the ratio R in xy plane at
rapidity  $Y$=0. The initial random distribution (panel a) do not
show any domain like structure with very high or low value of the
ratio. After the thermalisation of the fields at  T=200  MeV,  no
domain like structure is evident (panel b). Some small zones with
large   value  of  R  are  formed,  but  they  are  just  thermal
fluctuations. As will be shown below, thermal fields have  larger
correlation  than  random  fields  in quenched initial condition.
Small zones with large $r$ are manifestation of that correlation.
Large domain like structure with high/low value of  the  ratio  R
starts  to  emerge  after 10-15 fm of evolution and cooling. With
time domain  like  structure  grow.  At  other  rapidities  also,
similar  behaviour  is seen. Fig.2, clearly demonstrate that with
appropriate cooling law, multiple disoriented  chiral  condensate
domains  with  different  isospin  orientations  can  form as the
chirally symmetric phase  roll  back  to  broken  phase.  However
domain  formation  occur  quite  late  in the evolution. With the
scaling cooling law, by the time domain like  structures  emerge,
the  system  is cooled to $\sim$20 MeV. It is doubtful whether in
heavy ion collision, system can be allowed to be cooled  to  this
extent.  Current  wisdom  is  that  the hadrons freeze-out around
100-160 MeV. With this reservation in mind, it may be  said  that
even with thermal fields domains of disoriented chiral condensate
can form.

Above  behaviour is confirmed from the correlation study also. We
define a correlation function at rapidity $Y$ as \cite{as94},

\begin{equation}
C(r,\tau)  =  \frac{  \sum_{i,j} \pi(i) \dot \pi(j)}{\sum_{i,j}
|\pi(i)| |\pi(j)|}
\end{equation}

\noindent  where  the sum is taken over those grid points i and j
such that the distance between i and j is r. In fig. 3,  we  have
shown  the evolution of the correlation function at rapidity Y=0.
Initially there is  no  correlation  length  beyond  the  lattice
spacing  of  1  fm.  After  thermalisation,  the  correlation  is
increased marginally. At the thermalisation, the  fields  are  at
the   minimum   of  the  potential,  consequently  develops  some
correlation. Correlation donot increase till 10 fm of  evolution.
At  later  time, long range correlation develops, increasing with
time. At 20 and 30 fm, pions  separated  by  large  distance  are
correlated.

If  a  single DCC domain is formed in heavy ion collision, it can
be easily detected. Probability to obtain a particular fraction f
of (say) neutral pion from a single domain is  $P(f)=1/2\sqrt{f}$
\cite{ra93}.  However, our simulation indicate that a few numbers
of  domains  with  different  isospin  orientations  are  formed.
Naturally  the  resultant  distribution  will not be $1/\sqrt{f}$
type. In fig.4, $f$ distribution for 100 events, at zero rapidity
are shown. Density of pions  at  rapidity  Y  was  calculated  by
integrating over the space-time as,

\begin{equation}
N_\pi(Y) = \int \pi^2 \tau d\tau dx dy
\end{equation}

If the thermalised fields are evolved upto $\tau$=10 fm, then the
$f$  distribution  is sharply peaked around the isospin symmetric
average value of 1/3. At early stage of evolution,  as  indicated
in  fig.2 and 3, long range correlations or domain like structure
are not developed in individual events. If the fields are evolved
upto 20 or 30 fm, then there is definite  domain  like  formation
and  also  long  range  correlation  is  developed, in individual
events. The $f$ distribution still has average /13, but the width
is increased. This indicate that large fluctuations in the  ratio
$r$  from  event  to  event  is  expected  when there is DCC like
phenomena. Experimentally, those  events  for  which  neutral  to
total   pion   ratio  shows  considerable  fluctuations  will  be
promising candidate for DCC events.

In  summary, we have studied the phase transition and disoriented
chiral condensate formation in linear sigma model in presence  of
background.  The  background  was  represented  by  a white noise
source i.e. a heatbath. Evolving the fields in contact  with  the
heat  bath  for  sufficiently  long  time  such  that  fields are
thermalised, it was shown the the model undergoes 2nd order phase
transition around $T_c$=120 MeV. It was also shown  that  if  the
thermalised fields are cooled down sufficiently rapidly, multiple
disoriented  chiral  condensate  domains  are  formed. Long range
correlation also develops. However,  even  with  sufficient  fast
cooling,   with   thermalised   fields,  domains  or  long  range
correlations develop quite late in the evolution (after 10-15  fm
of  cooling,  when  the fields are cooled to 20 MeV or so). It is
doubtful whether in heavy ion collision, the hadronic  fluid  can
be  allowed to evolve and cool for such long time. It is possible
that they undergoes freeze-out much  earlier.  Then,  disoriented
chiral  condensate domains will not be formed. We have also shown
that with multiple  domain  formation,  probability  distribution
donot  follow  the  $1/\sqrt{f}$  type  of  law,  rather, it is a
Gaussian with average at the isospin symmetric value of 1/3,  but
with increased fluctuations. It is suggested that one should look
into  events with large fluctuations of the neutral to pion ratio
around the isospin symmetric value of 1/3.

\acknowledgements
The   author  would like to thank K, Rajagopal and C. Gale and J. Randrup for
discussions. He also  gratefully  acknowledge  the  kind
hospitality of Centre for Theoretical  Physics,  MIT  and  McGill
University where part of the work was done.

\begin{figure}
\centerline{\psfig{figure=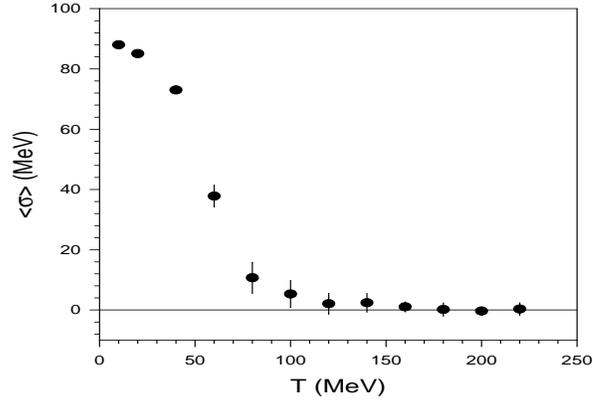,height=10cm,width=10cm}}
\vspace{-2cm}
\caption{Equilibrium  value  of the order parameter as a function
of the temperature.}
\end{figure}
\begin{figure}
\centerline{\psfig{figure=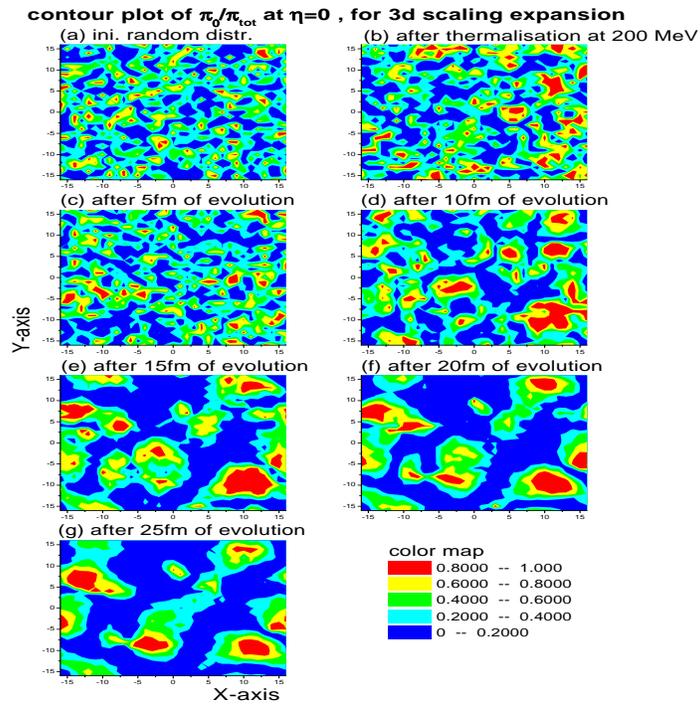,height=10cm,width=10cm}}
\caption{contour plot of the neutral to total pion ratio at rapidity Y=0 at different
stages of evolution.}
\end{figure}
\begin{figure}
\centerline{\psfig{figure=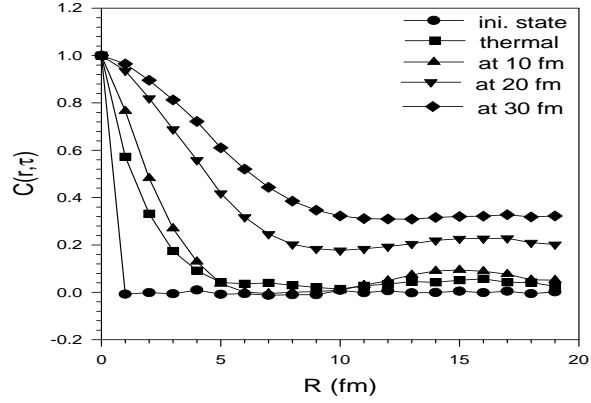,height=10cm,width=10cm}}
\vspace{-2cm}
\caption{Evolution of the correlation function at rapidty Y=0.}
\end{figure}
\begin{figure}
\centerline{\psfig{figure=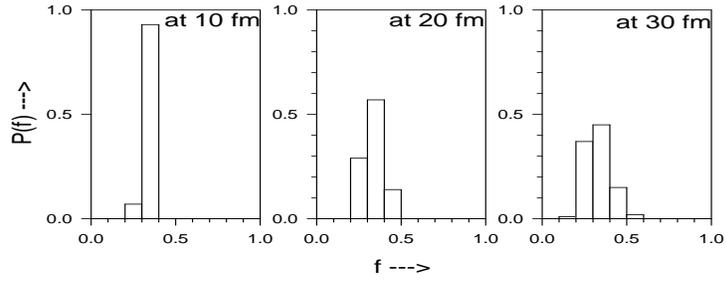,height=10cm,width=10cm}}
\vspace{-3cm}
\caption{Probabilty distribution for the neutral to total pion ratio at rapidity Y=0,
 for 100 events at different stages of evolution}
\end{figure}
\end{document}